\documentclass{article}
\usepackage[utf8]{inputenc}

\title{Loss and Recovery of Genetic Diversity in Adapting Populations of HIV}
\author{Pleuni S. Pennings \textsuperscript{1}, Sergey Kryazhimskiy \textsuperscript{2} , John Wakeley  \textsuperscript{3}}
\date{}

\usepackage{graphicx}
\usepackage{cite}
\makeatletter
\renewcommand{\@biblabel}[1]{\quad#1.}
\makeatother

\begin{document}

\maketitle

1. Department of Biology, Stanford University, Stanford, California, USA
2. FAS Center for Systems Biology and Department of Organismic and Evolutionary Biology, Harvard University, Cambridge, Massachusetts, USA
3. Department of Organismic and Evolutionary Biology, Harvard University, Cambridge, Massachusetts, USA

Correspondence: Pleuni S. Pennings, pleuni@stanford.edu

\section* {Abstract}
The evolution of drug resistance in HIV occurs by the fixation of specific, well-known, drug-resistance mutations, but the underlying population genetic processes are not well understood. By analyzing within-patient longitudinal sequence data, we make four observations that shed a light on the underlying processes and allow us infer the short-term effective population size of the viral population in a patient. Our first observation is that the evolution of drug resistance usually occurs by the fixation of one drug-resistance mutation at a time, as opposed to several changes simultaneously. Second, we find that these fixation events are accompanied by a reduction in genetic diversity in the region surrounding the fixed drug-resistance mutation, due to the hitchhiking effect. Third, we observe that the fixation of drug-resistance mutations involves both hard and soft selective sweeps. In a hard sweep, a resistance mutation arises in a single viral particle and drives all linked mutations with it when it spreads in the viral population, which dramatically reduces genetic diversity. On the other hand, in a soft sweep, a resistance mutation occurs multiple times on different genetic backgrounds, and the reduction of diversity is weak. Using the frequency of occurrence of hard and soft sweeps we estimate the effective population size of HIV to be $1.5 \times 10^5$ ($95\%$ confidence interval $[0.8 \times 10^5, 4.8 \times 10^5]$). This number is much lower than the actual number of infected cells, but much larger than previous population size estimates based on synonymous diversity. We propose several explanations for the observed discrepancies. Finally, our fourth observation is that genetic diversity at non-synonymous sites recovers to its pre-fixation value within 18 months, whereas diversity at synonymous sites remains depressed after this time period. These results improve our understanding of HIV evolution and have potential implications for treatment strategies.

\section*{Author Summary}

It is well known that HIV can evolve to become drug resistant if it acquires specific drug-resistance mutations, but the underlying population genetic processes are not well understood. We found that the evolution of drug resistance in HIV populations within infected patients occurs by one mutation at a time (as opposed to multiple mutations simultaneously) and involves both hard and soft sweeps. In a hard sweep, a mutation originates in a single viral particle and then spreads to the entire viral population within the patient. As this mutation increases in frequency, other mutations linked to it hitchhike to high frequencies, which greatly reduces genetic diversity in the population. In a soft sweep, on the other hand, the same resistance mutation originates multiple times on different genetic backgrounds, and hitchhiking may have very little or no effect on diversity. The fact that drug resistance evolves by means of both hard and soft sweeps implies that the HIV populations are limited by the supply of resistance mutations. Using the frequency of hard and soft sweeps we obtain a point estimate of 150,000 for the effective population size of the virus, a number that is much higher than estimates based on diversity at neutral (synonymous) sites, but much lower than the actual number of HIV infected cells in a human patient. 

\section* {Introduction}

  Understanding the process of adaptation is one of the basic questions in evolutionary biology.
  The evolution of drug resistance in pathogens such as HIV is a prime example of adaptation and a major clinical and public health concern because it leads to treatment failures.  

  The likelihood that a population adapts in response to an environmental challenge, e.g., a viral population in a patient develops resistance in response to a drug treatment, depends, among other things, on the amount of genetic diversity that the population harbors \cite{Hayden2011, Lauring2013}.
  The amount of genetic diversity in a population depends in turn on the details of the adaptive process.
   In the classical model of adaptation, a beneficial mutation arises once in a single individual and, as this mutation increases in frequency in the population, other mutations ``hitchhike" with it to high frequency \cite{MaynardSmithAndHaigh74}.
   In this process, referred to as a ``hard selective sweep'', genetic diversity at linked sites is strongly reduced when the beneficial mutation fixes in the population.
   However, if the population is sufficiently large, beneficial mutations will occur frequently or will be present in multiple copies prior to the onset of selection. 
   Then, multiple beneficial alleles may replace the wildtype without leading to a significant reduction in diversity.
   This is called a ``soft selective sweep'' \cite{Pennings2006a}.

   We use HIV as a model system to ask the following evolutionary questions. How much do selective sweeps affect genetic diversity at neighboring sites? How often does adaptation occur via soft and hard sweeps? What is the effective population size of HIV relevant for adaptation (later referred to simply as $N_e$)? How does diversity recover after a selective sweep at synonymous and non-synonymous sites?
   Three factors make HIV well suited to address these questions: (a) HIV evolves rapidly due to its short replication time and high mutation rate, (b) HIV populations in different patients act as independent replicates of the same evolutionary process, and (c) the main genetic targets of positive selection in HIV during antiviral treatment are known: these targets are drug-resistance mutations which have been well characterized. 

From the questions above, the estimation of the short-term effective population size ($N_e$) of HIV is perhaps the most interesting and clinically important. 
 We are interested in the  short-term effective population size because it is one of the determinants of the probability that drug resistance evolves in a certain generation. 	This effective population size is primarily determined by current processes (such as possibly background selection), and much less so by historical processes or events (such as bottlenecks).
   The question of the effective population size of HIV has been a subject of a heated debate because of enormous discrepancies in the estimates obtained by different methods (see \cite{KouyosAlthausBonhoeffer2006} for a review). 
   In an influential 1997 paper, Leigh-Brown used genetic diversity at synonymous sites and estimated the effective size of the viral population to be on the order of $10^3$ \cite{LeighBrown1997}.
   Around the same time, it became clear that the number of active virus-producing cells in the body of an infected person was around $10^8$ \cite{Haase1994} which lead to a natural expectation that the viral population should be very large with almost deterministic evolutionary dynamics \cite{Coffin1995}.
   However, Frost et al found that frequencies of resistance mutations before the start of treatment varied greatly between patients, which led them to conclude that stochastic effects play an important role in HIV, and the population size must therefore be $10^6$ or smaller \cite{Frost2000JIV}. 
   With more data and a more sophisticated linkage disequilibrium-based method Rouzine and Coffin inferred that the population size of the virus must be at least $10^5$ \cite{Rouzine1999}.
   Finally, a recent study by Maldarelli et al \cite{Maldarelli2013} places $N_e$ between $10^3$ and $10^5$, based on changes in allele frequencies over time.

   We estimate $N_e$ of HIV using previously published longitudinal sequence data from several HIV-infected patients under drug treatment \cite{Bacheler2000}.
   Our approach to estimating $N_e$ is most closely related to the recent estimates done in {\em Drosophila }\cite{Karasov2010}, {\em Plasmodium }\cite{Nair2007}, and a chimeric simian-human immunodeficiency virus \cite{Boltz2012}, but differs from these studies in that we use data from several patients for increased statistical power.
   The main idea of our method is as follows.
   Since the mode of adaptation (hard versus soft sweeps) depends on the supply of beneficial mutations which in turn depends on the effective size of the population, it is possible to use the observed frequency of hard and soft sweeps to estimate $N_e$.
   The key problem is to distinguish hard and soft sweeps in the sequence data.
   We address this problem by taking advantage of a unique property of codon 103 in the reverse-transcriptase gene, namely that there are two distinct nucleotide changes that convert the (wildtype) amino acid lysine into asparagine, which confers drug resistance. If both of these nucleotide changes are observed in a single sample, a soft sweep must have occurred. 

   While the effects of selection and demographic processes on the reduction of diversity have been extensively studied, the question of how diversity recovers after perturbations has received little attention. 
   Yet, understanding the dynamics of recovery is important because it determines how long signatures of selective sweep or bottlenecks remain detectable in genetic data. 
   Failure to account for the time it takes for diversity to reach its equilibrium after a perturbation can lead to biases in estimation of population genetic parameters such as $N_e$ \cite{KouyosAlthausBonhoeffer2006}.
   Using longitudinal data we show here that diversity after a selective sweep recovers faster at non-synonymous sites than at synonymous sites.
 
\section*{Results and Discussion}

\subsection*{Selective sweeps reduce genetic diversity}

  We study how the genetic composition of HIV populations in patients treated with a combination of reverse transcriptase (RT) and protease inhibitors changed over the course of about one year (see Ref. \cite{Bacheler2000} and {\bf Loss of diversity} for details on the data). Out of 118 patients studied in Ref. \cite{Bacheler2000}, we selected a subset of 30 patients in which the virus had no resistance mutations at the start of the study but fixed a resistance mutation at one of the later time points. All our analyses were done on this subset of patients (see \textbf{Loss of diversity} for details).  We note that in 24 out of 30 patients ($80\%$) we detect fixation of a single resistance mutation within the observation period, even though all patients were treated with multiple drugs (see \textbf{Suppl Table S1}). 

  Samples taken prior to treatment in these patients show nucleotide diversities of $\sim3\%$ at synonymous sites and $<1\%$ at non-synonymous sites.     
  At the time points when we detect fixations of drug-resistance alleles at one or several known sites we observe a substantial decrease in the levels of polymorphism in the 1Kb region that was sequenced (see Figures 1, 2, 3, and {\bf Loss of diversity}). 
  Between the last sample before and the first sample after the fixation of the first drug-resistance mutation, viral populations lose $53\%$ of their genetic diversity, measured by the median drop in per-site heterozygosity. This difference is significant, with $P<10^{-3}$ (all P-values we report are from Mann-Whitney tests). We observe no loss of diversity in control intervals in which no resistance mutation was fixed (see \textbf{Loss of diversity}).  
  
  The loss of diversity that we observe is a result of hitchhiking: when an adaptive mutation rapidly increases in frequency, it takes with it the genetic background on which it arose \cite{MaynardSmithAndHaigh74}.  
  Three factors can attenuate the observed loss of diversity after a sweep. 
  First, recombination can preserve some diversity by allowing sites at some distance from the selected site to escape the effects of the sweep \cite{KaplanEtAl89}. 
  Recombinational escape does not appear to be a factor in these data however because we do not observe a correlation between post-sweep heterozygosity and distance from the selected site (see \textbf{Recombination}).
  This is consistent with previously reported observations in HIV \cite{Nijhuis1998}.

  Second, the amount of diversity lost depends on the origin of the adaptive allele.
  Diversity may survive the fixation of an adaptive allele even in the absence of recombination if the selective sweep is ``soft'', i.e., if the same adaptive allele arises via multiple mutations on different genetic backgrounds \cite{Pennings2006b}.
  On the other hand, genetic diversity at neighboring sites is reduced dramatically if the selective sweep is ``hard'', i.e., if the adaptive allele originates through a single mutation on one genetic background \cite{MaynardSmithAndHaigh74}. 
  We demonstrate below that both soft and hard sweeps occurred in these patients.
  Importantly, whether soft or hard sweeps predominate in a population depends on the supply of adaptive mutations which is determined by the product of the beneficial mutation rate and the current effective population size.
  If adaptive mutations appear in the population on average more than once per viral generation, they are likely to fix via soft sweeps.
  If adaptive mutations appear on average less than once per generation, they typically fix via hard sweeps \cite{Pennings2006a}. 
  Thus, we can assess the supply rate of resistance mutations by determining whether the observed selective sweeps are soft or hard.

  Third, the observed loss of diversity depends on whether new mutations occur between the fixation of a resistance allele and the time when samples are taken and the sweep is observed. 
  Since we do not know the actual time of fixation, we do not know the absolute amount of diversity loss due to the sweep.
   However, this fact does not obscure informative patterns of relative loss of diversity, for synonymous versus non-synonymous mutations and for soft versus hard sweeps.

\subsection*{Estimating $N_e$ from the frequency of  soft sweeps}

  In 17 of the 30 HIV populations ($57\%$), we observe the fixation of just one, well-known drug resistance mutation, K103N  (lysine to asparagine) in RT. This mutation confers high-level resistance to the non-nucleoside RT inhibitor used by the patients \cite{Bacheler2000}. 
  Two different nucleotide changes produce the same K103N amino acid change.
  The wildtype codon AAA encoding lysine can mutate to either codon AAC or codon AAT both encoding asparagine which confers resistance.
  This property is unique to the K103N mutation among all resistance mutations observed in these patients (see \textbf{Suppl. Table S1}) and can be used to distinguish soft and potentially hard sweeps. 
  
  In 7 of these 17 populations ($41\%$) both codons (AAC and AAT) that code for asparagine are present in the population in the first sample after the sweep (Figure 1). 
  These are certain to be soft sweeps. As expected, the median reduction of diversity in these populations is only $15\%$ and not significantly different from 0. 
  In the remaining 10 patients ($59\%$) only one codon is present after the sweep (AAC in 6 and AAT in 4 patients; Figure 2).
  These sweeps could be either soft or hard. 
  If in these 10 ``one-codon'' patients the resistance allele typically arose via the same number of unique mutation events as in the 7 ``two-codon'' patients, we would expect to see a similarly small reduction in diversity at linked sites in these two groups of patients. 
  In contrast, we find that in the 10 ``one-codon'' viral populations, genetic diversity is reduced by $71\%$ ($P= 0.003$), and the difference between the ``two-codon'' and ``one-codon'' populations is significant ($P=0.007$).
  It is possible that two codons initially fixed in the ``one-codon'' patients but one of the codons was lost (for example, due to another sweep) before the sample was taken.
  This scenario would have been plausible if the time intervals between samples were, for some reason, longer in the ``one-codon'' than in the ``two-codon'' patients.
  However, we observe no statistical difference in the inter-sample time intervals between the ``two-codon'' and ``one-codon'' patients ($P=0.48$).
  We therefore conclude that the selective sweeps in the ``one-codon'' patients are due to fewer mutational origins than in the ``two-codon'' patients, and some are likely hard sweeps (see Figure 2).
  
  The fact that we observe both soft and hard sweeps implies that the K103N mutation appears roughly once per generation \cite{Pennings2006a}.
  A more accurate estimate of the mutation supply rate $\theta/2$ based on the frequencies of AAC and AAT alleles in the post-sweep samples yields $\theta/2 = 0.3$ new K103N mutations per generation ($95\%$ confidence interval $[0.17, 0.97]$). 
  Assuming that we sample 6 viral isolates per patient (which is the median in our dataset), theory predicts that, if $\theta/2 = 0.3$, $30\%$ of populations would show a hard sweep signature (i.e., the sweep would be due to a single origin), which is consistent with our observations.
  Assuming the point mutation rate of $2\times 10^{-6}$ per generation for transversions \cite{Abram2010}, we estimate the effective population size of HIV populations to be $N_e = 1.5\times10^5$ $[0.8\times 10^5,4.8\times 10^5]$ (see {\bf Estimating the supply rate of resistance mutations} for details). 
 
We estimated $N_e$ based on data from the 17 patients in which the K103N mutation fixed. Although we cannot do the same analysis with the other resistance mutations in the dataset, we find that the median reduction in diversity in the other 13 patients is intermediate between the ``one-codon'' and ``two-codon'' K103N patients (see Suppl. Figure S1), which suggests that K103N mutation is not special and that similar processes are taking place at the other resistance sites. 
 Further, the haplotype structures in two patients with amino-acid changes in the 190th codon position suggest a soft sweep in one and a hard sweep in the other (see Suppl. Figure S2).  Thus adaptation at the 103rd codon position appears to be representative of adaptation at other codon positions.

 There are other amino acid changes that are similar to the lysine to asparagine (K to N) change in that they can be achieved by two different nucleotide changes. 
 We provide a list of these amino acid changes (limited to those changes that are one mutational step away) in Suppl. Table S2. One of the changes on this list is the methionine to leucine (M to L) change, which is relevant for drug resistance in HIV. The M46L change in the Protease gene leads to resistance to certain Protease Inhibitors. A six-patient dataset from 1997 \cite{Zhang1997} provides two examples of the fixation of the M46L change from wildtype (ATG) to resistant (TTG or CTG). In one of these two patients, we observe a mixture of the TTG and CTG codons (thus a soft sweep) whereas in the other patient we only observe the TTG codon (potentially a hard sweep, see Suppl. Figure S3). Of the other four patients, two did not fix the M46L mutation and two fixed it at the same time as another resistance mutation. It is difficult to draw conclusions based solely on the data from this study because it consists of only 6 patients \cite{Zhang1997}, but they do provide independent support for our observation that drug resistance mutations in HIV fix by means of both soft and hard sweeps. 

 An interesting question is whether the resistance mutations that fix stem from standing genetic variation or are due to de novo mutations during treatment. If the resistance mutations stem from standing genetic variation, then our estimate of $N_e$ would reflect the pre-treatment effective population size, whereas, if the mutations would be due to de novo mutation, our estimate would reflect the effective population size during treatment.
  We compared the times of occurrence of soft and hard sweeps relative to the onset of treatment and found that these were statistically indistinguishable.
  There are three possible explanations for this observation.
  First, almost all observed sweeps are due to standing genetic variation, and our estimate of $N_e$ reflects the pre-treatment population.
  Second, almost all sweeps are due to new mutations and our estimate reflects only the on-treatment $N_e$.
  Third, sweeps we observe originated from both standing genetic variation and de novo mutation, but the start of the treatment had no discernible effect on the effective population size, and our estimate reflects both the pre-treatment and on-treatment population size.

  We currently have no statistical power to distinguish between these three possibilities, and so it is possible that, if the second explanation is true, the pre-treatment population size is much larger than our estimate.
  However, some of the sweeps we observed occurred very early during treatment (within one or two months), and it seems likely that at least these early sweeps are due to standing genetic variation (see also \cite{Pennings2012}).  
  Note that, counter to intuition, whether mutations stem from the standing genetic variation or not does not affect the expected number of origins in a sample (see Suppl. Figure S4 and \cite{Pennings2006a}).

Our estimate of $N_e$ is similar to earlier estimates that were also based on resistance mutations \cite{Frost2000JIV} and other beneficial mutations \cite{Rouzine1999}. Our approach may be viewed as a combination of these two previous approaches. Rouzine and Coffin \cite{Rouzine1999} tested whether a recombinant or double mutant genotype is present, given that two beneficial mutations are sweeping through the population at the same time.  Similar to a second mutation that leads to a soft sweep, a double mutant or recombinant can only occur in a population with sufficiently high mutational or recombinational input, so that the frequency of such a genotype in a sample permits the estimation of the mutational or recombinational input.  
  Frost et al \cite{Frost2000JIV} used inferred pre-treatment frequencies of two resistance mutations at the same codon (RT M184I and M184V) to estimate the effective population size. They found that the two mutations were present at different pre-treatment frequencies in seven patients. In a very large population, the frequency of such mutations would be close to the expected frequency ($\mu/s$) in every patient, but in smaller populations frequencies vary stochastically, and the spread in the observed frequencies can be used to obtain an upper bound for the population size. 

Our estimate of the supply rate of resistance mutations is much lower than would be expected if the effective population size were equal to the number of virus-infected cells in the body, which is estimated to be $10^8$ \cite{Haase1994,Coffin1995}. 
At the same time, it is much higher than estimates of the effective population size based on the level of diversity at synonymous sites (assuming neutrality of synonymous variation) which would be around $10^3$ \cite{LeighBrown1997}.
 Next we discuss possible reasons for these discrepancies.

\subsection*{Why is our estimate higher than $N_e$ from synonymous diversity?}

We propose three possible explanations for the large difference between our estimate based on the frequency of soft sweeps and the estimate based on the traditional analyses of synonymous diversity. All three explanations may be contributing to this difference.
First, some synonymous mutations may be deleterious 
\cite{Sanjuan2011}. 
 Second, positive and negative selection on linked sites (draft and background selection) reduce diversity at synonymous sites, but may have a weaker effect on beneficial mutations such as K103N. Our observation that the fixation of drug resistance mutations is accompanied by a reduction in diversity provides evidence for this explanation.
 Third, synonymous diversity may take a long time to recover after a sweep or a bottleneck.
 We discuss the third option in more detail and provide evidence from the data for this explanation.

  We observe that diversity at both synonymous and non-synonymous sites steadily recovers after the sweep (Figure 3). 
  However, even after 6 to 18 months, synonymous diversity remains significantly depressed, with a median reduction of $-53\%$ ($P=0.01$) compared to pre-sweep diversity, up from $-66\%$ ($P=10^{-4}$) directly after the sweep. 
  In contrast, after 6 to 18 months after the sweep, non-synonymous diversity is fully recovered, with the median amount of diversity increased by $+5\%$ (n.s.)\ compared to pre-sweep, up from $-32\%$ ($P=0.01$) directly after the sweep.
  Hence, even relatively infrequent selective sweeps can keep synonymous diversity at a level much below the equilibrium.

  The observation that stronger negative selection leads to faster recovery is consistent with recent predictions about the approach of the distribution of allele frequencies to stationarity at a single locus \cite{SongAndSteinrucken2012}.
  This effect is well known in systems biology: stronger negative feedback leads to a faster the recovery time \cite{Alon2006}. 
  A heuristic analysis ({\bf Text S1}), similar to analyses in \cite{malecot1969mathematics, Nei1975, Gordo2005}, shows that neutral sites recover half of their diversity in roughly $N_e$ generations while deleterious sites recover half of their diversity in roughly $1/s$ generations, where $s$ is the strength of selection against deleterious mutations. Interestingly, for realistic values of the mutation rate, the recovery time is independent of the mutation rate (see {\bf Text S1 and Suppl. Figure S5}). 
  
  The dynamics of how diversity recovers are likely not specific to HIV but typical to any large population.
  In humans, for example, Europeans have reduced synonymous diversity, but a similar amount of non-synonymous diversity, compared to Africans \cite{Lohmueller2008,Kiezun2013}.
  This is consistent with the explanation that non-synonymous diversity has had enough time to recover since the out-of-Africa bottleneck, but synonymous diversity has not.
  This observation leads to a counterintuitive practical implication.
  Since non-synonymous diversity recovers faster, it is usually closer to its equilibrium than synonymous diversity.
  Thus non-synonymous sites may in fact be more useful for classical population genetics inference than synonymous sites.

\subsection*{Why is our estimate lower than the number of infected cells?}

  HIV populations consist of a heterogeneous collection of genotypes, many of which may carry deleterious and advantageous mutations. 
  A new mutation arises in one such genotype and is therefore linked, at least temporarily, to other mutations already present in it.
  The probability that a new mutation eventually fixes or goes extinct thus depends not only on its own selective effect but also on the combined fitness of the genetic background in which it arose \cite{DesaiFisher2007Genetics,Schiffels2011}.
  The effects of linkage on the fates of new mutations are complex and comprise an active area of research \cite{DesaiFisher2007Genetics,Neher2011,Schiffels2011,Walczak2012,Desai2012,Rouzine2010, Rouzine2003,Rouzine2008,Neher2010Genetics,Good2012}, but the qualitative picture is straightforward.
  The fates of neutral, deleterious and weakly beneficial mutations are entirely determined by the background in which they arise \cite{Schiffels2011}:  only mutations that arise on high-fitness genotypes have a chance to persist in the population (Figure 4a, b). 
   The number of segregating neutral mutations in such a population will be small because high-fitness individuals comprise only a small fraction of the population \cite{DesaiFisher2007Genetics,Walczak2012,Desai2012}.
  At the other extreme, adaptive mutations with very large selective effects survive and spread irrespective of the genetic background in which they arise (Figure 4c).
  The effective population size for such mutations may be close to the census population size \cite{Karasov2010}.
  The effective population size for mutations with intermediate effects is somewhere in between (Figure 4d).
  
  The effective size we estimate from the K103N mutation is several orders of magnitude smaller than the estimated census size.
  This could mean that the distribution of fitnesses in the viral population due to ongoing positive and negative selection is wide, so that even a highly beneficial mutation such as K103N needs to appear in relatively fit genetic background to have a good chance of survival.
  The width of the distribution of fitnesses in within-patient HIV populations is currently unknown.
  Theory shows that the width of the fitness distribution depends on the rates and typical effects of beneficial and deleterious mutations and on the viral census size \cite{DesaiFisher2007Genetics,Good2012,Goyal2012}, quantities that are difficult to estimate.
  Neher and Leitner recently estimated that at least $15\%$ of non-synonymous mutations have a selection coefficient exceeding $0.8\%$ \cite{Neher2010}.
    In another study, Batorsky et al concluded that the rate of adaptation of HIV rate is consistent with an exponential distribution of fitness effects of adaptive mutations with mean $0.5\%$ \cite{Batorsky2011}.
    If these estimates are accurate, they could imply that the distributions of fitnesses in within-patient HIV populations are narrow so that fixation of highly beneficial mutations like K103N should not be affected by linked selection.

  Other effects may help explain why the effective size of HIV is smaller than the census size.
  For example, the treatment itself may reduce the population size of the virus before resistance evolves. If this is true, then soft sweeps must be especially common in patients who were treated with inferior drugs in the 1980s and 1990s, because early treatments were not as successful in reducing the population size of the virus. If treatment reduces the effective population size, then this would also mean that soft sweeps should be more common in patients who fail early versus patients who fail later (after more time on treatment). We find no evidence for such an effect in the current data set (results not shown). 
  Alternatively, spatial population structure and epistatic effects may reduce the effective population size for resistance mutations.
 
  We obtained our population size estimate under the assumption---most certainly violated in reality at least to some extent---that all patients have the same constant effective population size.
  Unfortunately, we do not have sufficient statistical power to relax this assumption.
  Longitudinal data with denser and deeper sampling is required to study inter-patient and temporal heterogeneity in viral adaptation. 

\subsection*{Conclusions}

  We observe soft and hard sweeps in HIV, which leads to an estimated effective population size relevant for adaptation of around $1.5 \times 10^5$. This number is much higher than what the observed level of synonymous diversity suggests. 
  An important caveat of our approach is that we assume that the supply rate of beneficial mutations is similar in all patients, which may not be true. Larger samples are needed to estimate the patient-specific supply rate of adaptive mutations. 
  Our results show that neither synonymous diversity, nor the census population size provide sufficient information to infer a population's adaptive potential, which explains, in part, why it is hard to predict whether and when adaptive evolution will occur.  
  Our finding that adaptive evolution in HIV is limited by the supply rate of beneficial mutations suggests that a relatively modest reduction of the effective population size may help prevent the evolution of drug resistance, and it is possible that the reduction of $N_e$ that results from modern combination therapy, does just that \cite{Pennings2012}. 
 In addition, if the adaptive potential of the virus is limited because a large proportion of the beneficial mutations land on inferior genetic backgrounds, then the use of mutagenic drugs, in combination with standard antiretroviral drugs, should  reduce the adaptive potential of the virus \cite{Mullins2011}.   
    Finally, our results confirm what has already been predicted theoretically, that the idea of a single effective population that can be used to describe different aspects of the behavior of a population does not work 
  \cite{Karasov2010, Weissman2012}.

  \section*{Material and Methods}
  
  \subsection*{Loss of diversity}
Out of a larger dataset \cite{Bacheler2000}, we used viral sequences from patients for which we had samples at two consecutive time-points that satisfy the following two criteria. (1) At the first time-point, no known resistance mutations to any of the drugs used in the trial was present at more than $30\%$ frequency in the sample. See below for the list of drugs and mutations. (2) At the next time-point, at least one drug resistant allele increases in frequency by at least $70\%$. There were 30 such patients. In most cases (26 out of 30) the frequency of the mutations changes from 0 to $100\%$.
  The four exceptions are patient 89 (mutation {\bf G190S} changed from $0$ to $75\%$), patient 168 (mutation {\bf K103N} changed from $0$ to $83\%$), patient 22  (mutation {\bf K103N} changed from $134$ to $100\%$, while mutation {\bf V82A} changed from $0$ to $100\%$,), and patient 81 (mutation {\bf K103N} changed from $14$ to $100\%$).
  In most cases (24 out of 30) only a single drug-resistance mutation went to fixation (2 mutations in patients 22, 56, 87, 91, 154 and 166). For details per patient, see supplementary table S1. 

The patients were treated with Zidovudine, Lamivudine, Efavirenz and Indinavir \cite{Bacheler2000}. There are many known mutations that confer resistance to one or more of these drugs. We are interested in the fixation of the first resistance mutation in a viral population. We used the following list of major drug-resistance mutations (the number is the codon and the letter is the amino-acid that confers resistance). Protease: 46IL, 82AFT, 84V; Reverse Transcriptase: 41L, 62V, 65R, 67N, 70R, 75I, 77L, 100I, 101P, 103N, 106MA, 108I, 116Y, 151M, 181CI, 184VI, 188LCH, 190SA, 210W, 215YF, 219QE, 225H \cite{Johnson2010}.

Our dataset consists of coding sequences of the regions of the Pol gene that encode for protease (all 297 basepairs) and reverse transcriptase (first 689 basepairs). 
We analyze separately all third codon position sites and all first and second codon position sites. 
Most observed mutations at third codon position sites have no effect on the amino acid, and we expect these synonymous mutations to be neutral or nearly neutral. Throughout the paper, we refer to mutations at first and second position sites as non-synonymous and mutations at third position sites as synonymous. 
Most mutations at the first or second codon positions change the amino acid, and we expect that most such non-synonymous mutations are selected against. 

No change in genetic diversity was observed in 27 control intervals that also started with drug susceptible virus but in which no fixation of drug resistant amino-acids occurred (largest frequency change of drug resistant amino-acid in these intervals was $30\%$).

We found no difference in pre-sweep diversity at third codon position sites between hard sweep and soft sweep patients. 

\subsection*{Recombination}
  To determine whether recombination affects the reduction of diversity, we split the sequences into a part close to the selected site (less than 50, 100, or 200 basepairs distance from the selected site) and the complementary part far from the selected site.
  We found no difference in the amount of diversity loss in the close versus far sites.
  Earlier work suggests that selection may be so strong, relative to the recombination rate, that the sweeps can affect the entire genome \cite{Nijhuis1998, Neher2010}. 

\subsection*{Estimating the supply rate of resistance mutations}
  We can use the counts of the two beneficial alleles (AAT and AAC at the 103rd amino acid of RT) in the post-sweep samples to estimate the the supply rate of resistance mutations.
  We assume that the mutation rate $\mu_\mathrm{AAT}$ to the AAT codon and the mutation rate $\mu_\mathrm{AAC}$ to the AAC codon are equal to each other, $\mu_\mathrm{AAT} = \mu_\mathrm{AAC} \equiv \mu $.
  If $N_e$ is the effective population size, then $\theta = 2N_e\mu$, and $\theta/2$ is the total population-wide beneficial mutation rate, which we also call ``the supply rate of resistance mutations''.

  Theory predicts that the population frequency $x$ of the AAC allele follows a beta-distribution with parameters $(\theta/2, \theta/2)$ (see \cite{Pennings2006a}).
  The number $k$ of AAC alleles in a sample of size $n$ follows the binomial distribution with parameters $n$ and $x$.
  Thus, the likelihood of observing $k_1, \dots, k_M$ AAC alleles in samples of size $n_1, \dots, n_M$ in $M$ patients is given by the product of beta-binomial distributions,
  \begin{equation}
    L(k_1, \dots, k_M; \theta) = \prod_{i=1}^M {n_i \choose k_i}
    \frac{B(k_i + \theta/2, n_i-k_i+\theta/2)  }{B(\theta/2, \theta/2)},
    \label{eq: likelihood}
  \end{equation}
where $B(a,b)$ is the beta-function.
  By maximizing expression (\ref{eq: likelihood}), we obtain the maximum likelihood estimate $\hat \theta = 0.62$.
  The estimated supply rate of K103N mutations is $\hat\theta/2 = 0.31$. 
  The $95\%$ bootstrap confidence interval for the supply rate obtained by resampling patients with replacement is 
  $[0.17,0.97]$. 

  We estimate the effective population size as $\hat \theta/2\mu$.
  Because we only consider the K103N mutation, $\mu \approx 2 \times 10^{-6}$ \cite{Mansky1995, Abram2010}.
  Note that this number is lower than the typical per site mutation rate, because most ($\sim 85\%$) mutations in HIV are transitions, whereas the A to C and A to T mutations that create the K103N change, are both transversions. We thus estimate the effective population size to be $1.5 \times 10^5$ ($95\%$ confidence interval $[0.8 \times 10^5, 4.8 \times 10^5]$).

For a given sample size and a given supply rate of beneficial mutations, we can also predict the probability that the beneficial mutation originates from 1, 2 or more mutational origins following 
\cite{Pennings2006a}. For a sample of size 6 (the median in the dataset) and the estimated supply rate of beneficial mutations of $0.31$, the predicted probability that all observed beneficial alleles in the sample stem from a single origin is $30\%$, the probability that they stem from 2 origins is $43\%$, 3 origins: $22\%$ and 4, 5 or 6 origins: $6\%$. This shows that even if the supply rate is exactly the same in all populations, the observed sweep signature can vary widely. 
  
\section*{Acknowledgements}
  The authors thank 
  Graham Coop, 
  Michael Desai, 
  David Enard, 
  Nandita Garud,
  Ben Good, 
  Joachim Hermisson, 
  Rajiv McCoy, 
  Richard Neher,
  Dmitri Petrov,
  Ariel Weinberger, 
  Ben Wilson,  
  and two anonymous reviewers for comments and discussions.

\bibliography{references}

\pagebreak

\begin{figure}[h!]
\centering
\includegraphics[scale=1]{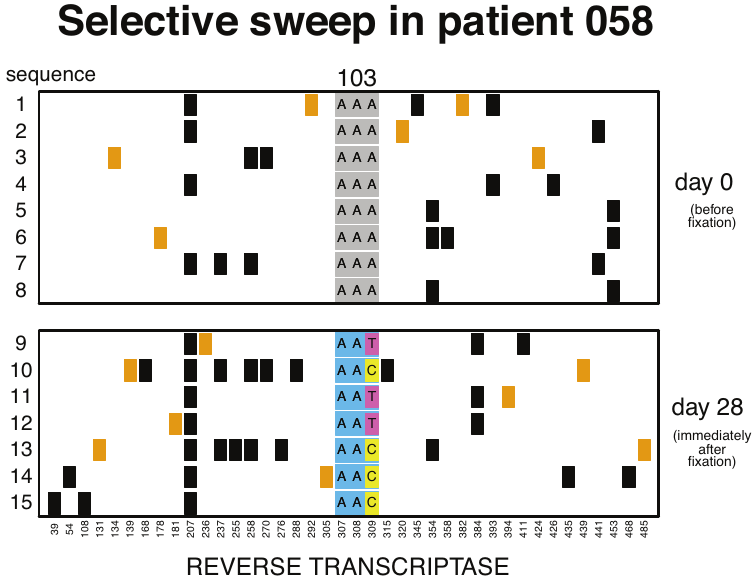}

{\bf Figure 1. Soft sweep in patient 058.} Codon AAA coding for lysine at position 103 was replaced by day 28 with a mixture of codons AAT and AAC both coding for asparagine which confers resistance to NNRTI drugs.
  Genetic diversity close to the selected site was not significantly reduced.
  The plot shows only the polymorphic sites among the first 500 basepairs of the reverse transcriptase region.
  Each row represents a sequenced viral isolate.
  Each column represents a polymorphic site, with the derived synonymous and non-synonymous polymorphisms shown in black and orange respectively.
  Codon 103 is shown explicitly. It is grey when coding for lysine (susceptible) and blue when coding for asparagine (resistant). The responsible mutations are colored pink (T) and yellow (C). 
\label{SoftSweep}
\end{figure}

\newpage

\begin{figure}[h!]
\centering
\includegraphics[scale=1]{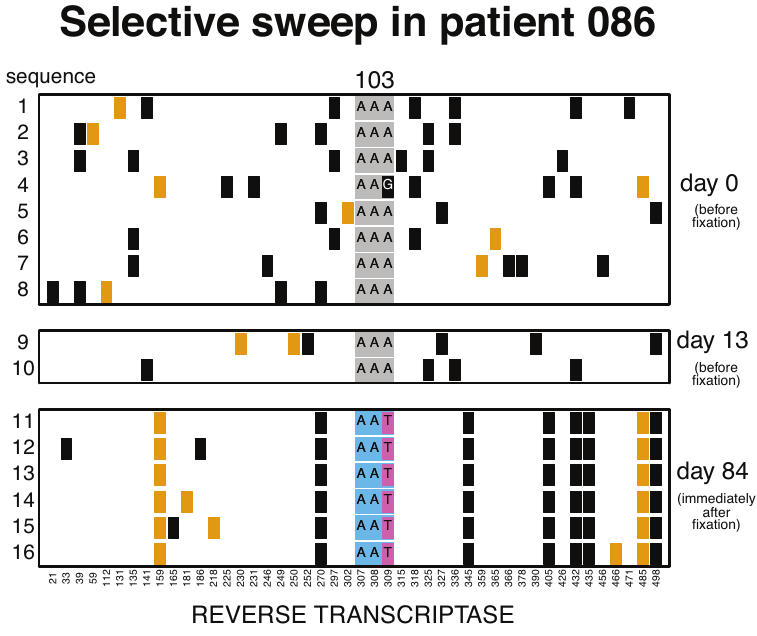}

{\bf Figure 2. Putatively hard sweep in patient 086.}
  Codon AAA coding at position 103 was replaced by day 84 with codon AAT.
  Genetic diversity around the selected site was strongly reduced.
  Notations as in Figure 1.
\label{HardSweep}
\end{figure}

\newpage

\begin{figure}[h!]
\centering
\includegraphics[scale=1]{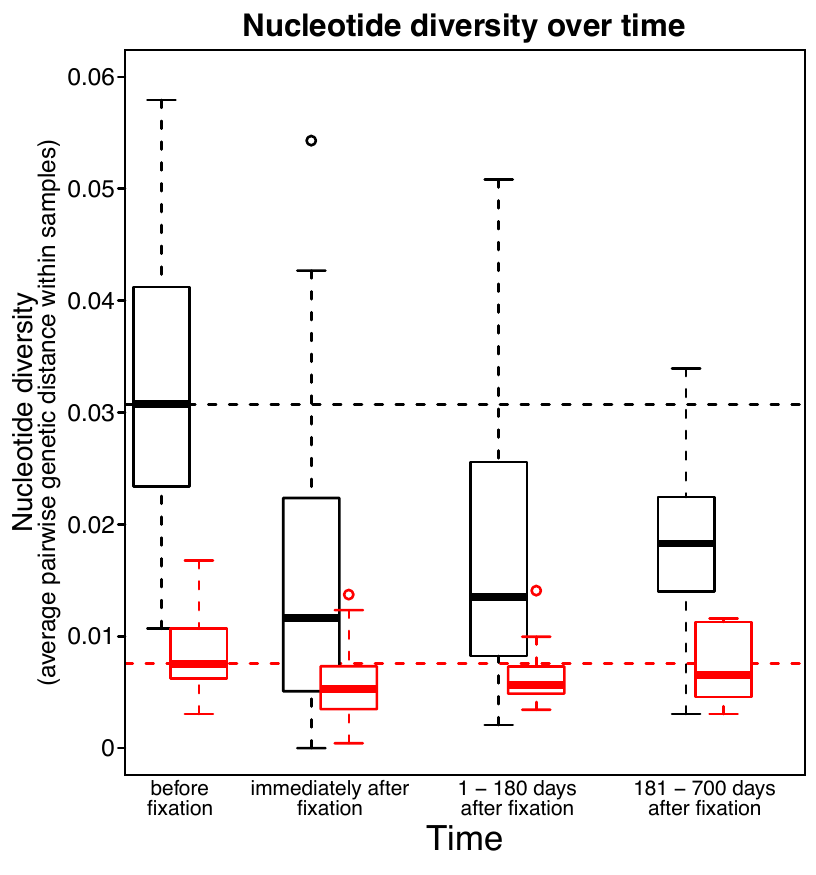}

{\bf Figure 3. Nucleotide diversity over time in 30 patients.} Nucleotide diversity at synonymous (black) and non-synonymous (red) sites.
  The time point ``before fixation'' is the last sample in each patient taken before the observed fixation of a resistance mutation.
  The time point ``immediately after fixation'' is the first sample in each patient in which the drug resistance mutation is observed to be fixed.
  The third and the fourth time points denote samples in each patient taken 1--180 days and 181--700 days after the observed fixation of the resistance mutation respectively.
  Dashed horizontal lines show the nucleotide diversity at synonymous (black) and non-synonymous (red) sites at the time point ``before fixation'' to provide a reference for subsequent time points. 
\label{RecoveryFigure}
\end{figure}

\newpage

\begin{figure}[h!]
\centering
\includegraphics[scale=1]{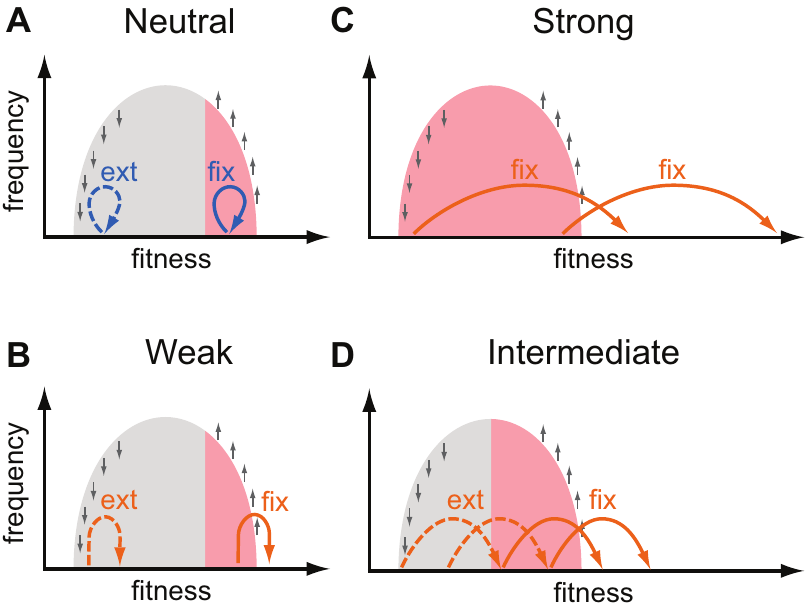}

{\bf Figure 4. Fates of mutations in a heterogeneous population.}
  This schematic shows the fates (``fix'', fixation; ``ext'' extinction) of mutations with different selective effects, as indicated in each panel, depending on the fitness of the genotype they occur in.
  The light gray area shows the distribution of fitnesses in the population.
  Small gray arrows show how natural selection changes the distribution of genotypes: fitter than average genotypes increase in frequency, less fit than average genotypes decrease in frequency.
  Colored arrows show the effect size of new mutations: neutral mutations (effect size $0$) are shown in blue (panel a.), adaptive mutations (effect size $>0$) are shown in orange (panels b. weakly beneficial, panel c. strongly beneficial, panel d. intermediate beneficial effect).
  Mutations that are destined for fixation (extinction) are shown with solid (dashed) arrows.
  Mutations with a large beneficial effect can fix if they arise in any background (panel c), whereas mutations with a small beneficial effect can fix only if they arise in a very fit background (panels a., b., and d.).
  The resulting effective population size for mutations with a given effect size is shown as red shading.
 \label{WaveMuts}
\end{figure}

\newpage

\begin{figure}[h!]
{\bf Supplementary Text S1. TextS1.pdf Recovery of diversity}. In this section, we analyze how genetic diversity recovers after a selective sweep in the data and in a mathematical model. 
\end{figure}

\subsection*{Recovery of diversity}
Because the patients were followed for approximately one year (median, appr. 200 generations \cite{Fu2001MBE}), it is possible to determine whether the viral populations in the patients recover from the observed selective sweep. We binned the observations in three bins. Bin 1: directly after the fixation event (the day the resistance mutation was detected, most likely this is shortly after it was fixed and allowed an increase in viral load). Bin 2: observations between 1 and 180 days after the fixation was observed. Bin 3: between 181 and 700 days after the fixation was observed. If a second drug resistance mutation became fixed in the virus of a patient the sample in which this was observed and any later samples were removed from the analysis. 

In order to model the recovery of heterozygosity, we assume that during the periods between sweeps of resistance mutations there is no positive or balancing selection, only mutation, random drift, and negative selection.  We further assume that negative selection on linked sites simply reduces the effective population size and may thus be captured by the random drift term in  standard, single-locus population-genetic models \cite{Nagylaki92}. 

We assume two alleles: the `non-mutant' $A_1$ and the `mutant' $A_2$, with relative fitnesses 1 and $1-s$, respectively.  With probability $u$ $A_1$ mutates to $A_2$, and with probability $v$ $A_2$ mutates to $A_1$.  Reproduction occurs according to the haploid Wright-Fisher model, with non-overlapping generations and constant population size $N$.  If the current frequency of the mutant is $x$, then 
\begin{eqnarray}
x^{\prime\prime} &=& \frac{x(1-s)(1-v) + (1-x) u}{1-sx}  
\end{eqnarray}
gives the frequency after selection and mutation.  The number of mutants in the next generation is binomial with parameters $N$ and $x^{\prime\prime}$, so that its frequency $X$ has expectation $E[X] = x^{\prime\prime}$ and variance $\textrm{Var}[X] = x^{\prime\prime} (1 - x^{\prime\prime})/N$.

We are interested in the recovery of heterozygosity, and so consider
\begin{eqnarray}
E[\Delta H] &=& E[2X(1-X) - 2x(1-x)] \\ [7pt]
&=& 2 x^{\prime\prime} (1-x^{\prime\prime}) \left( 1 - \frac{1}{N} \right) - 2x(1-x). 
\end{eqnarray}
where the $(1 - \frac{1}{N})$ term reflects the loss of heterozygosity due to identity by descent (coalescence). 
We seek a simple heuristic formula that will aid in understanding the recovery of heterozygosity after a selective sweep.  Assuming that $s$, $u$, $v$, and $1/N$ are all small, which is appropriate for HIV, and further assuming that the mutant frequency $x$ is small, gives 
\begin{eqnarray}
E[\Delta H] &\approx& 2 u - \left( s + 3u + v + \frac{1}{N} \right) H ,  
\end{eqnarray}
in which $H = 2 x (1-x) \approx 2 x$.  We apply this result to the recovery of heterozygosity at synonymous and non-synonymous sites heuristically using a continuous-time approximation. 

For synonymous sites, we assume that all mutations are neutral and that $u,v\ll\frac{1}{N}$.  Then $\frac{dH}{dt} = 2 u - \frac{1}{N} H$, and if $H(0) = 0$, we have $H(t) = 2Nu(1-\exp(-t/N))$.
The response time, $t_{half}$, defined as the time it takes for $H$ to recover $50\%$ of its loss of diversity, can be found by solving $0.5 = \exp(-t_{half}/N)$ giving $t_{half}=N\log 2$. Note that under the assumption $u,v\ll\frac{1}{N}$, the response time is independent of the mutation rates and it is possible to estimate $N$ independently from $u$.  
For non-synonymous sites, we assume that selection is stronger than both random drift and mutation, or and solve $\frac{dH}{dt} = 2 u - s H$ to obtain $H(t) = 2(u/s)(1-\exp(- s t))$.
In this case the response time is $t_{half}=s^{-1} \log 2$.  Thus $s$ may be estimated independently from $u$.  These expressions for $t_{half}$ illustrate that non-synonymous sites will recover faster than synonymous sites if $s > \frac{1}{N}$.  Suppl Figure 2 shows that these approximate, heuristic expressions agree well with simulations as long as the mutation rate is sufficiently small.  

Our conclusions are in agreement with previous results by Gordo and Disonisio \cite{Gordo2005} who described that more deleterious mutations approach mutation-selection equilibrium faster than less deleterious mutations. Song and Steinr\"{u}cken \cite{SongAndSteinrucken2012} have recently described a method for studying the approach to stationarity of the entire distribution of allele frequencies, and also illustrated that recovery is faster when mutations are more deleterious.

\begin{figure}[h!]

\centering
\includegraphics[scale=.5]{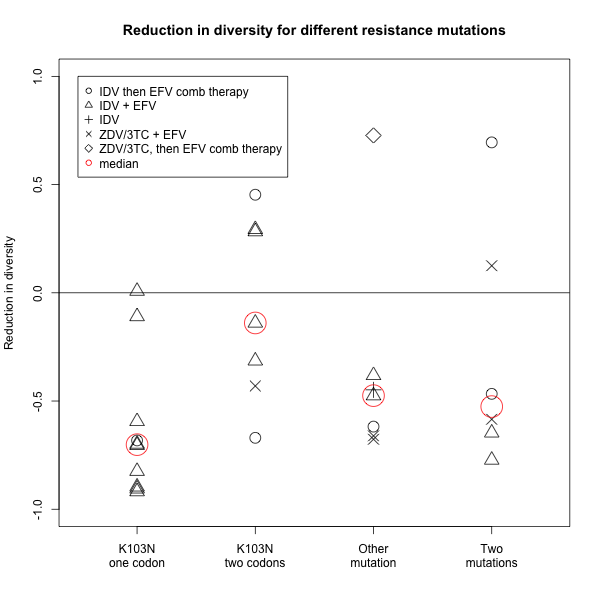}

{\bf Supplementary Figure S1. Nucleotide diversity reduction due to different fixations.} Nucleotide diversity at all sites. 
The patients are split in four groups: the first group in which one codon of the K103N was fixed (K103N one codon), a group in which a mixture of two codons of K103N was fixed (K103N two codons), a group in which a different amino acid change was fixed (Other mutation) and a group in which two resistance mutations were fixed (Two mutations). Large red circles show the median for each group. 
The shape of the points reflects the treatment the patient was on:
circle:  treated with indinavir then later switched to efavirenz combination therapy, 
triangle: treated with indinavir + efavirenz, 
plus-sign: treated with indinavir,  
cross:  treated with ZDV/3TC + efavirenz   
diamond: treated with ZDV/3TC, later switched to efavirenz combination therapy. 
\label{Pat159}
\end{figure}

\begin{figure}[h!]
\centering
\includegraphics[scale=1]{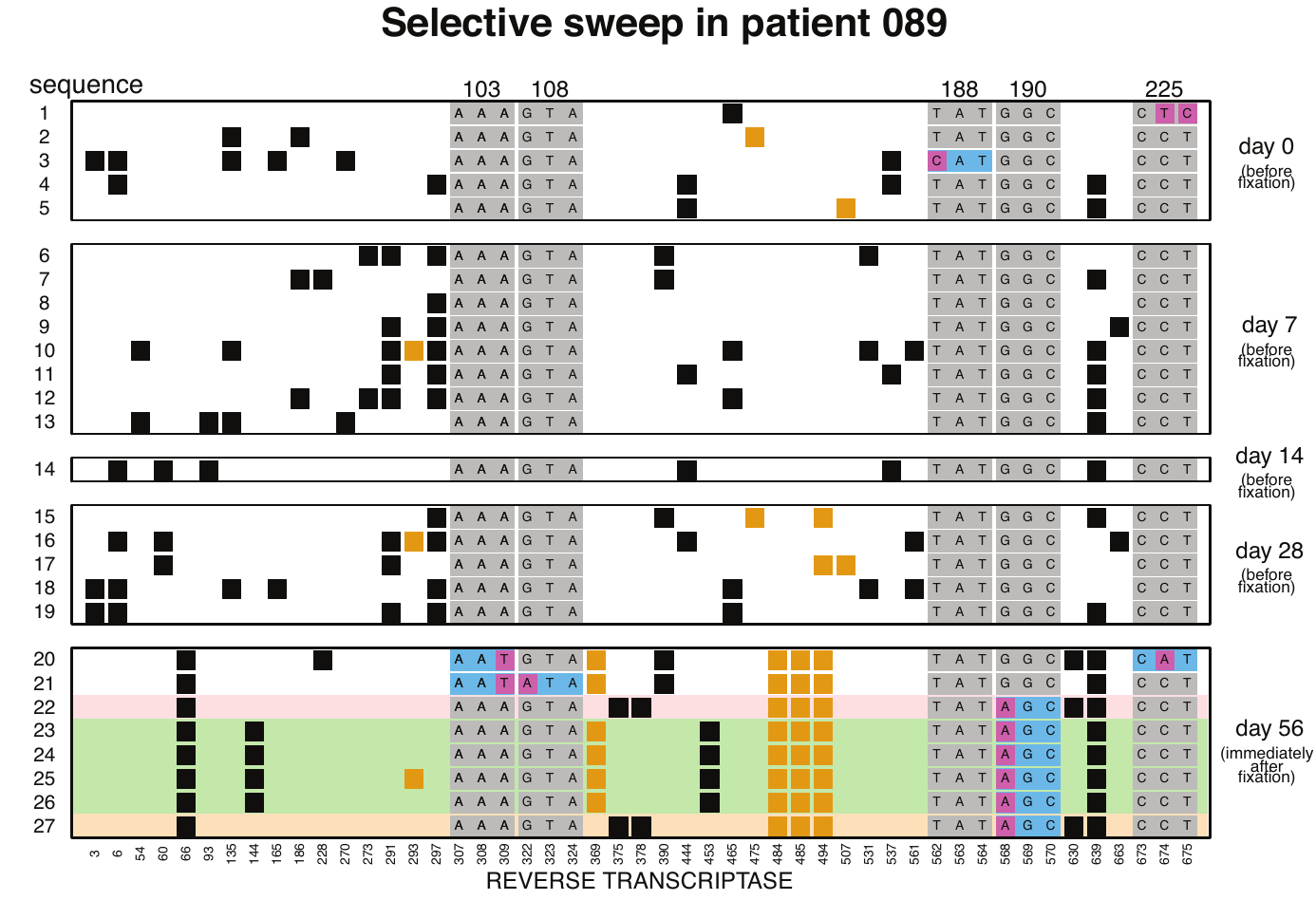}

{\bf Supplementary Figure S2. Soft and hard sweep example.}
A. Soft sweep in patient 089. Codon GGC coding for glysine at position 190 was replaced by day 56 by codon AGC coding for serine, which confers resistance to NNRTI drugs. The G190S mutation appears to have occurred on at least two different backgrounds. As predicted by theory \cite{Pennings2006b}, the site is embedded in a region of high linkage disequilibrium. The different haplotypic backgrounds are indicated by differently colored pastel backgrounds. Genetic diversity was reduced by $38\%$.
  The plot shows the polymorphic sites in the reverse transcriptase region, excluding all singletons.
  Each row represents a sequenced viral isolate.
  Each column represents a polymorphic site, with the derived synonymous and non-synonymous polymorphisms shown in black and orange respectively.
  Codons 103, 108, 188, 190 and 225 are all linked to NNRTI resistance and are shown explicitly. 
  They are grey when when in the susceptible state and blue when in the resistant state. Mutations in these codons are  colored pink. 
  \end{figure}

\begin{figure}[h!]

\centering

  \includegraphics[scale=1]{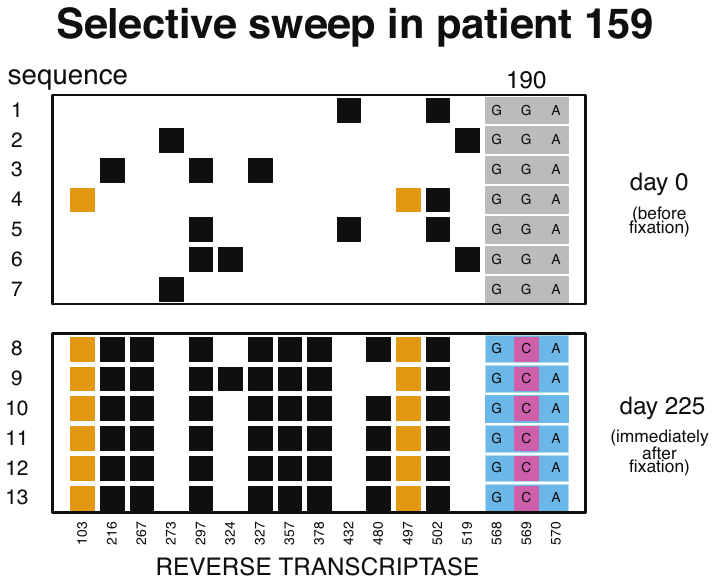}
  
{\bf Supplementary Figure S2. Soft and hard sweep example.}
  B. Hard sweep in patient 159. 
Codon GGA coding for glysine at position 190 was replaced by day 225 by codon GCA coding for alanine, which confers resistance to NNRTI drugs. The G190A mutation appears to have occurred on at least a single haplotypic background. Genetic diversity was reduced by $67\%$.
  The plot shows the polymorphic sites in the reverse transcriptase region, excluding all singletons.
  Each row represents a sequenced viral isolate.
  Each column represents a polymorphic site, with the derived synonymous and non-synonymous polymorphisms shown in black and orange respectively.
  Codons 190 is shown explicitly.  It is colored grey when when in the susceptible state and blue when in the resistant state. Mutations in this codon are colored pink. 
\label{Pat089}
\end{figure}

\begin{figure}[h!]
\centering
\includegraphics[scale=1]{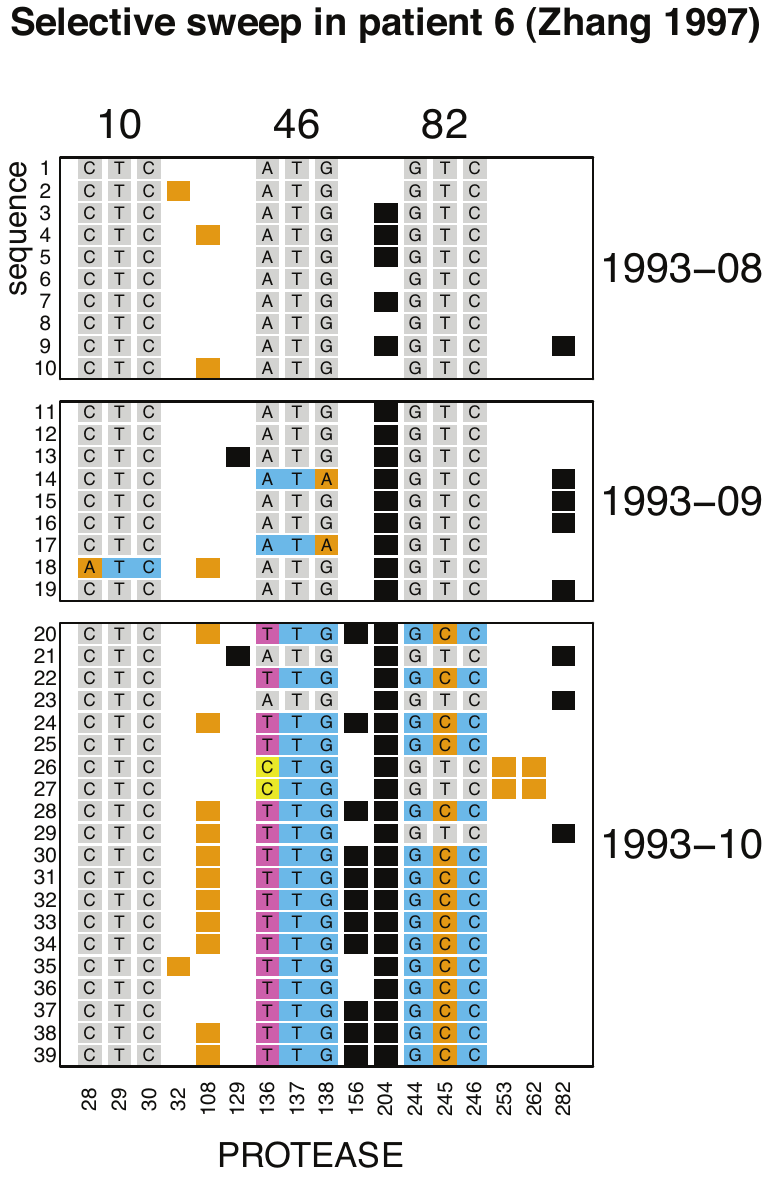}

{\bf Supplementary Figure S3. Soft and hard sweep example.}
A. Soft sweep in patient 6 \cite{Zhang1997} 
Codon ATG coding for methionine at position 46 was replaced in 1993/10 by codons CTG and TTG, both coding for leucine, which confers resistance to Protease inhibitor drugs. 
  The plot shows the polymorphic sites in the protease region, excluding all singletons.
  Each row represents a sequenced viral isolate.
  Each column represents a polymorphic site, with the derived synonymous and non-synonymous polymorphisms shown in black and orange respectively.
  Codons 10, 46, 71 are linked to Protease Inhibitor drug resistance and shown explicitly.  They are colored grey when when in the susceptible state and blue when in the resistant state. Mutations in these codons are  colored pink. 
  The viral load in this patient was around $10^6$ before treatment, it went down approximately $1.5$ log values and rebounded quickly to its original level (see figure 1 in \cite{Zhang1997}).
  
    \end{figure}

\begin{figure}[h!]

\centering

    \includegraphics[scale=1]{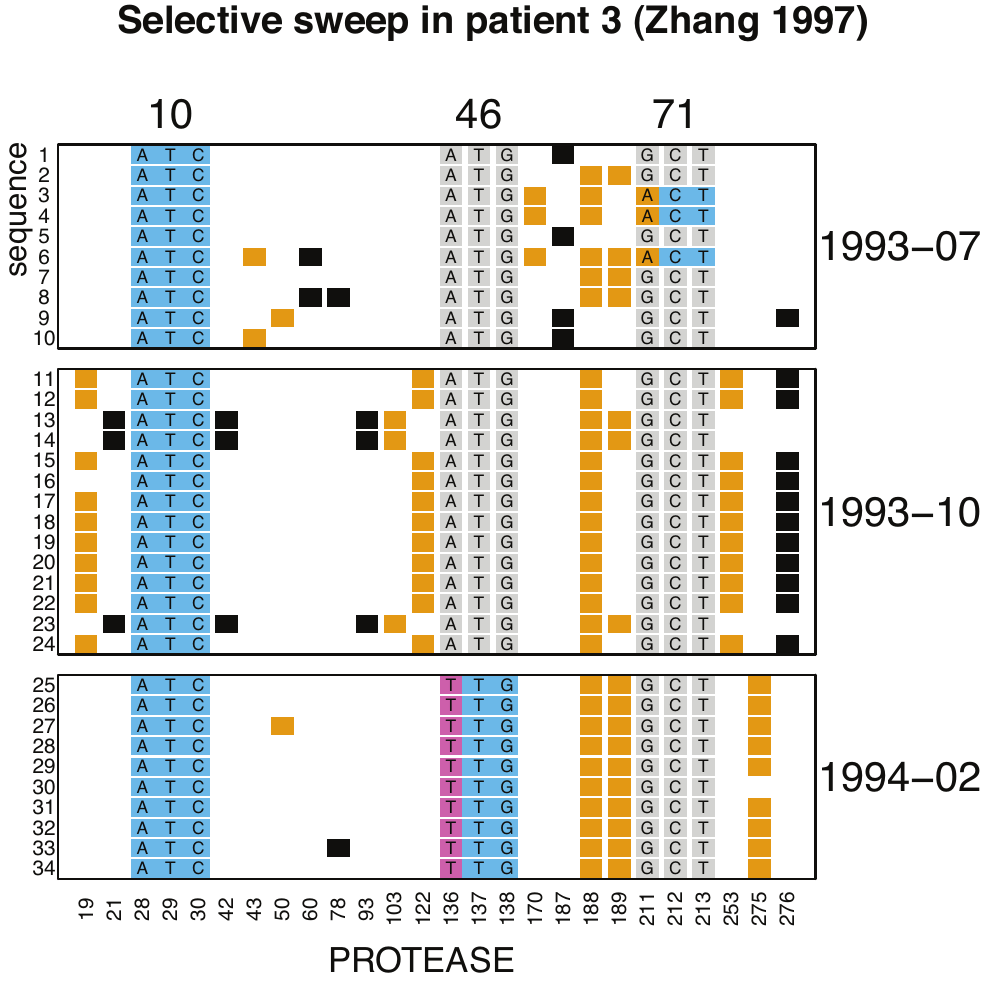}

{\bf Supplementary Figure S3. Soft and hard sweep example.}
  B. Hard sweep in patient 3 \cite{Zhang1997}
Codon ATG coding for methionine at position 46 was replaced in 1994/02 by codon TTG, which codes for leucine, which confers resistance to Protease inhibitor drugs. 
  The plot shows the polymorphic sites in the protease region, excluding all singletons.
  Each row represents a sequenced viral isolate.
  Each column represents a polymorphic site, with the derived synonymous and non-synonymous polymorphisms shown in black and orange respectively. 
  Codons 10, 46, 71 are linked to Protease Inhibitor drug resistance and shown explicitly.
   They are colored grey when when in the susceptible state and blue when in the resistant state. Mutations in these codons are  colored pink. 
  The viral load in this patient was slightly higher than $10^5$, it went down to below the detection limit of $10^4$ and stayed low for more than six months, after which it rebounded (see figure 1 in \cite{Zhang1997}).
\label{Pat6}
\end{figure}

\begin{figure}[h!]
\centering
\includegraphics[scale=.6]{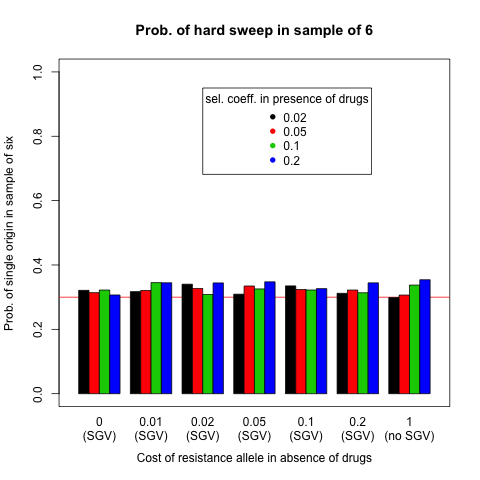}

{\bf Supplementary Figure S4. Probability of a single origin (hard sweep) in a sample of size 6.} Simulation results show that the number of origins is independent of whether or not standing genetic variation is involved, and if standing genetic variation is involved, the number of origins is independent of the selection coefficient of the allele before or after the start of treatment. The horizontal line denotes the theoretical expectation \cite{Pennings2006a}. The bars denote the probability that a sample of size six consists of just one origin, averaged over 1000 runs. The population size in the simulations is $150,000$ and the mutation rate $2\times10^{-6}$, so that $\theta/2 = 0.3$. 
\label{X}
\end{figure}

\begin{figure}[h!]
\centering
\includegraphics[scale=.5]{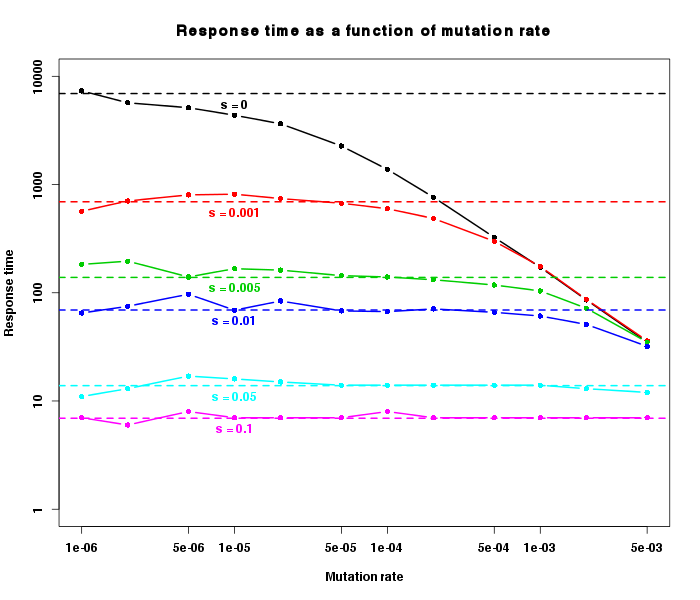}

{\bf Supplementary Figure S5. Predicted response time.} Comparison of the response time $t_{half}$, defined as the time it takes for $H$ to recover $50\%$ of its loss of diversity, from the heuristic model and from simulations. Dashed lines show $t_{half}=s^{-1}\log(2)$ (for $s>0$) or $t_{half}=N\log(2)$ (for $s=0$).  Dots display averages over 1000 simulations of a haploid Wright-Fisher population with $N=10^4$ individuals for each pair of values of $u$ and $s$.  In all cases, $v=u$. 

\label{Thalf}
\end{figure}

\begin{table}[ht]
{\bf Supplementary Table S1.} Patient details. For each of the (a.) 30 patients that were included in the analysis, this table gives (b.) the codon(s) of the resistance mutation(s) that occurred (with nucleotide changes indicated for all mutations except the 103rd codon of RT, P82 indicates the 82nd codon in Protease, all other codons are in RT), (c.) the level of diversity before and (d.) after the sweep, (e.) the sample size of the first sample after the sweep, (f.) the number of AAC and (g.) AAT alleles observed in the first sample after the sweep, and (h.) the treatment the patient was receiving.

\begin{tabular}{ |p{.6cm}|p{3.5cm}|p{.8cm}|p{.8cm}|p{.6cm}|p{.6cm}|p{.6cm}|p{4.7cm}|}
  \hline
 a. pat ient& b. resistance codon / mutation & c. div pre & d. div after & e. sam. size & f. 103 AAC  & g. 103 AAT  & h. treatment \\ 
  \hline
  5 & 103 & 0.018 & 0.006 & 4  & 3 & 1 & IDV then EFV comb therapy \\ 
   7 & 103 & 0.008 & 0.008 & 6  & 6 & 0 & IDV + EFV \\ 
  12 & 103 & 0.024 & 0.016 & 2  & 1 & 1 & IDV + EFV \\ 
  13 & 103 & 0.018 & 0.002 & 7  & 0 & 7 & IDV + EFV \\ 
  20 & P82 (GTC $\rightarrow$ ACC) & 0.017 & 0.010 & 6 & 0 & 0 & IDV \\ 
  22 & 103, P82 (GTC $\rightarrow$ GCC) & 0.022 & 0.005 & 4  & 4 & 0 & IDV + EFV \\ 
  24 & 103 & 0.012 & 0.017 & 7  & 5 & 2 & IDV then EFV comb therapy \\ 
  32 & 103 & 0.012 & 0.004 & 8  & 8 & 0 & IDV + EFV \\ 
  44 & 103 & 0.019 & 0.017 & 3  & 3 & 0 & IDV + EFV \\ 
  56 & 103, 62 (GGC$\rightarrow$ GTC) & 0.012 & 0.004 & 2  & 2 & 0 & IDV + EFV \\ 
  57 & 103 & 0.010 & 0.013 & 6  & 3 & 3 & IDV + EFV \\ 
  58 & 103 & 0.012 & 0.016 & 7  & 4 & 3 & IDV + EFV \\ 
  63 & 103 & 0.012 & 0.005 & 7  & 7 & 0 & IDV + EFV \\ 
  66 & 103 & 0.016 & 0.003 & 5  & 5 & 0 & IDV + EFV \\ 
  70 & P82 (GTC $\rightarrow$ GCC) & 0.017 & 0.007 & 6 & 0 & 0 & IDV then EFV comb therapy \\ 
  71 & 103 & 0.018 & 0.005 & 7  & 0 & 7 & IDV + EFV \\ 
  77 & 103 & 0.025 & 0.002 & 7  & 0 & 7 & IDV + EFV \\ 
   81 & 103 & 0.012 & 0.010 & 8  & 7 & 1 & IDV + EFV \\ 
   83 & 190 (GGA $\rightarrow$ GCA)& 0.015 & 0.008 & 8 & 0 & 0 & IDV + EFV \\ 
   86 & 103 & 0.029 & 0.003 & 6  & 0 & 6 & IDV + EFV \\ 
   87 & 103, P82 (GTC $\rightarrow$ GCC) & 0.011 & 0.006 & 8  & 7 & 1 & IDV then EFV comb therapy \\ 
  89 & 190  (GGA $\rightarrow$ AGC) & 0.020 & 0.012 & 8 & 0 & 0 & IDV + EFV \\ 
  91 & 103, P82  (GTC $\rightarrow$ GCC) & 0.012 & 0.020 & 2  & 2 & 0 & IDV then EFV comb therapy \\ 
   95 & 103 & 0.012 & 0.004 & 6  & 6 & 0 & IDV then  EFV comb therapy \\ 
  154 & 103, 100 (TTA $\rightarrow$ ATA) & 0.020 & 0.008 & 7  & 7 & 0 & ZDV/3TC + EFV \\ 
  159 & 190 (GGA $\rightarrow$ GCA) & 0.015 & 0.005 & 6 & 0 & 0 & ZDV/3TC + EFV \\ 
  166 & 103, 184 (ATG $\rightarrow$ GTG) & 0.009 & 0.011 & 4  & 4 & 0 & ZDV/3TC + EFV \\ 
  167 & 184  (ATG $\rightarrow$ GTG)& 0.015 & 0.026 & 7 & 0 & 0 & ZDV/3TC,  then EFV comb therapy \\ 
  168 & 103 & 0.024 & 0.014 & 6  & 2 & 3 & ZDV/3TC + EFV \\ 
  171 & 184 (ATG $\rightarrow$ GTG)& 0.024 & 0.008 & 2 & 0 & 0 & ZDV/3TC + EFV \\ 
   \hline
\end{tabular}

\label{SupplementaryTable S1}
\end{table}

\begin{table}[ht]
\small
{\bf Supplementary Table S2.} List of amino acid changes that can be achieved by two or more single nucleotide changes and thus allow the observation of soft sweeps at the beneficial site itself. In almost all cases, the relevant mutations are transversions (TV). In the last three cases (marked with an asterisk), one of the mutations is a transition (TS). In these cases, the difference in mutation rates (higher for transitions) will make it less likely to observe a soft sweep.
\begin{tabular}{ |p{1.cm}|p{1.cm}|p{1.cm}|p{5cm}|p{3cm}|}
  \hline
 a. From amino acid& b. From codon & c. \break To amino acid & d. \break To codons & e. \break Relevance for HIV drug resistance \\ 
  \hline
  K &AAA & N & AAC (TV), AAT (TV) & relevant for RT K103N\\ 
M &ATG & L & CTG (TV), TTG (TV) & relevant for RT M41L and PRO M46L\\ 
C &TGC & S & AGC (TV), TCC (TV)& \\ 
C &TGT & S & AGT (TV), TCT (TV)& \\ 
D &GAC & E & GAA (TV), GAG (TV)& \\ 
D &GAT & E & GAA (TV), GAG (TV)& \\ 
E &GAA & D & GAC (TV), GAT (TV)& \\ 
E &GAG & D & GAC (TV), GAT (TV)& \\ 
G &GGA & R & AGA (TS), CGA (TV)& \\ 
G &GGG & R & AGG (TS), CGG (TV)& \\ 
H &CAC & Q & CAA (TV), CAG (TV)& \\ 
H &CAT & Q & CAA (TV), CAG (TV)& \\ 
I &ATA & L & CTA (TV), TTA (TV)& \\ 
K &AAG & N & AAC (TV), AAT (TV)& \\ 
L &TTA & F & TTC (TV), TTT (TV)& \\ 
L &TTG & F & TTC (TV), TTT (TV)& \\ 
N &AAC & K & AAA (TV), AAG (TV)& \\ 
N &AAT & K & AAA (TV), AAG (TV)& \\ 
Q &CAA & H & CAC (TV), CAT (TV)& \\ 
Q &CAG & H & CAC (TV), CAT (TV)& \\ 
R &AGA & S & AGC (TV), AGT (TV)& \\ 
R &AGG & S & AGC (TV), AGT (TV)& \\ 
T &ACC & S & TCC (TV), AGC (TV)& \\ 
T &ACT & S & TCT (TV), AGT (TV)& \\ 
V &GTA & L & CTA (TV), TTA (TV)& \\ 
V &GTG & L & CTG (TV), TTG (TV)& \\ 
W &TGG & C & TGC (TV), TGT (TV)& \\ 
W &TGG & R & AGG (TV), CGG (TS)& \\ 
S &AGC & R & CGC (TV), AGA (TV), AGG (TV)& \\ 
S &AGT & R & CGT (TV), AGA (TV), AGG (TV)& \\ 
F &TTC & L & CTC (TS), TTA (TV), TTG (TV)& *\\ 
F &TTT & L & CTT (TS), TTA (TV), TTG (TV)& *\\ 
M &ATG & I & ATA (TS), ATC (TV), ATT (TV)& *\\ 
\hline
\end{tabular}

\label{SupplementaryTable S1}
\end{table}

\end{document}